\documentclass[hyper]{JHEP3}

\input{epsf}
\usepackage{epsfig}
\usepackage{amssymb}
\usepackage{amsfonts}
\usepackage{amsbsy}
\usepackage[all]{xy}
\usepackage{amsmath}

\usepackage{amssymb,amscd}
\usepackage{mathrsfs}
\usepackage{amsmath,amsthm}
\def\be{ \begin{equation} }
\def\ee{ \end{equation}}

\def\det{{\rm det}}

\def\exp{{\rm exp}}

\def\Hom{{\rm Hom}}
\def\I{{\rm i}}

\def\log{{\rm log}}
\def\mod{{\rm mod}}

\def\half{\frac{1}{2}}
\def\p{\partial}

\def\one{{\hbox{ 1\kern-.8mm l}}}
\def\CD {{\cal D}}

\def\CF {{\cal F}}

\def\CH {{\cal H}}

\def\CK {{\cal K}}

\def\CV {{\cal V}}

\def\CH {{\cal H}}

\def\CS {{\cal S}}

\def\CU {{\cal U}}

\def\IA{\mathbb{A}}

\def\IC{\mathbb{C}}

\def\IR{{\mathbb{R}}}

\def\IZ{{\mathbb{Z}}}

\def\rmk#1{\bigskip\noindent{\bf Remarks} }
\title{A Comment On Berry Connections}

\author{Gregory W.~Moore  \\
  NHETC and
Department of Physics and Astronomy, Rutgers University \\
126 Frelinghuysen Rd., Piscataway NJ 08855, USA\\
{\tt gmoore@physics.rutgers.edu} }

\abstract{When families of quantum systems are equipped with a continuous family of Hamiltonians
such that there is a gap in the common spectrum one can define a notion of a Berry connection.
In this note we stress that, in general, since the Hilbert bundle defining the family of quantum
systems does not come with a canonical trivialization there is in fact not a single Berry connection
but rather a family of Berry connections. Two examples illustrate that this remark can have
physical consequences.
\today }

\begin{document}

\section{Introduction And Conclusion}

In this paper we comment on a little subtlety in the definition
of ``Berry phases'' and ``Berry connections'' which seems to have
been overlooked in reviews such as
\cite{Berry2,BohmBook,ChenThesis,Shapere:1989kp}. Put briefly:
The usual discussion explains that a Berry connection depends on
three pieces of data:

\begin{enumerate}

\item A bundle of Hilbert spaces $\pi:\CH \to X$ over a space $X$ of control parameters.

\item A family of Hamiltonians parametrized by $X$.

\item A choice of energy cutoff defining ``low energy states.''

\end{enumerate}

In fact, there is a fourth piece of data that is needed in the
construction: One must also choose a connection on the bundle $\CH$. In general
there are many choices for such a connection, and the choice
 can have physical consequences. What principles ought to  be
used to make this choice is an interesting question. It is part of the specification of
the physical problem, and a full discussion of such principles is far beyond the scope of this
letter.

In this letter the need to include the fourth piece of data is illustrated with two examples.
These examples are, unfortunately, not too dramatic, but they do suffice
to make the point that the subtlety under discussion can have physical
consequences.  The first example resolves a
minor paradox about the standard expression for electric polarizability per unit
volume due to valence electrons in an insulator \cite{KingSmithVanderbilt,RestaRev,RestaVanderbilt} and is
ultimately rather trivial as a physical effect. The other effect is more interesting and
is associated with the formula for the ``axion angle'' of an insulator in $3+1$ dimensions
\cite{Essin:2008rq,Essin:2010in,Qi:2008ew}. It leads to the $3+1$ dimensional version of
the quantum Hall effect, explored by B. Halerpin et. al. about 25 years ago
\cite{Halperin1,Halperin2}. (In fact, demanding mathematical naturalness in the formulation
of the Berry connection led me on an independent path to the $3+1$ dimensional QHE. But it was
then pointed out to me by A. Furusaki and N. Read that the $3+1$ dimensional QHE
is a standard and well-known result.)

While the two examples we give here are, perhaps, uninspiring, we believe that this subtlety
will play a role in many other physical examples. One such example arises in studying
families of two-dimensional conformal field theories with toroidal target spaces. These naturally
lead to a bundle of CFT statespaces over Narain moduli spaces. There is more than one
``natural'' connection on these statespaces so that any discussion of phenomena associated
to Berry connections on low-lying states (e.g. the moduli themselves) will be subject to
the subtlety we are discussing. Another example where we expect our considerations to play
a role is in higher-dimensional $tt^*$ geometry \cite{Cecotti:2013mba}.

There is also a derivation of ``the'' Berry connection from integrating out heavy modes (a.k.a. fast modes) in
a path integral \cite{Kuratsuji:1985kc,Moody:1989vh}. The subtlety we are discussing is related to the choice of
boundary conditions used for the heavy fields.
\footnote{I thank A. Kapustin for a discussion on this point.}
It might be interesting to explore this approach in more detail. There should also be
a parallel discussion for families of quantum systems with noncommutative
control parameters, as described in \cite{Moore:2017wlq}, and again we leave that
for the future.

\section*{Acknowledgements}

I would like to thank A. Abanov,
L. Fidkowski, D. Freed,  J. Fr\"ohlich, A. Furusaki, B. Halperin, A. Kapustin, A. Khan, T. Mainiero, L. Molenkamp, K. Rabe,
  N. Read, N. Seiberg, R. Shankar, D. Vanderbilt, and E. Witten for helpful discussions and correspondence.    Some of this paper was written while visiting the Institute for Advanced Study in Princeton and I am very
grateful to the IAS for hospitality and support from   the IBM Einstein Fellowship of the Institute for Advanced Study. Some of this work was done at the Aspen Center for Physics  (under NSF Grant No. PHY-1066293). I also acknowledge support by the DOE under grant
DOE-SC0010008 to Rutgers.

\section{Definition Of Berry Connections}

\subsection{Hilbert Bundles}

Let $X$ be a topological space. $X$  is to be thought of as a space of control parameters,
parametrizing some family of quantum systems. For simplicity we will take $X$ to be a connected space. Let
\be
\pi: \CH \to X
\ee
be a Hilbert bundle. This means the fibers $\pi^{-1}(x) := \CH_x$ are
Hilbert spaces and over a suitable open cover $\{\CU_{\alpha} \}$ there are isomorphisms
  $\varphi_\alpha: \pi^{-1}(\CU_{\alpha})\cong \CU_{\alpha} \times \CH_0$, where $\CH_0$ is a
  fixed Hilbert space,  and the transition functions $\varphi_{\beta}\circ\varphi_\alpha^{-1}$
  on patch overlaps $\CU_{\alpha\beta} = \CU_{\alpha} \cap \CU_{\beta}$ are  continuous maps
  to the unitary group of $\CH_0$, in a
suitable topology. (The details of the topology we use can be found in Appendix D
of \cite{Freed:2012uu}.)  A crucial point for this note will be the distinction
between a \emph{trivializable} Hilbert bundle and a \emph{trivial} Hilbert bundle.
The latter is a bundle which is literally a Cartesian product:
\be
\CH= X \times \CH_0
\ee
with $\pi$ the projection onto the first factor and $\CH_0$ is a fixed separable Hilbert
space, say $\IC^N$ for the finite-dimensional case and $\ell^2(\IZ)$ for the infinite-dimensional
case.  By contrast, a bundle $\pi: \CH \to X$ is
\emph{trivializable} if there is a bundle isomorphism to the trivial bundle $X \times \CH_0$.
What this means, in practice, is that for all $x\in X$ there is a basis $\{ \psi_{n,x} \}$ of
the Hilbert space $\CH_x$ that varies ``continuously'' (or ``smoothly'' if we want to differentiate)
as a function of $x$. By choosing some basepoint $x_0 \in X$ we can choose an isomorphism $\CH_{x_0}\cong \CH_0$
and then such a basis defines an isomorphism
\be
\Phi_{x_0, x}: \CH_x \rightarrow \CH_{x_0}.
\ee
A trivializable bundle together with a \underline{choice} of trivialization will be said to be
\emph{trivializ\underline{ed}}.

The distinction between trivializable and trivialized Hilbert bundles might seem like
hopelessly arcane solipsistic mathematical hair-splitting to most physicists, but we will see that
it can be important. A relevant preliminary remark is that if there are   different trivializations of $\CH$
corresponding to choices of bases   $\{ \psi_{n,x} \}$ and $\{ \tilde \psi_{n,x} \}$ then
$\tilde \Phi_{x_0,x} \circ \Phi_{x_0,x}^{-1}: X \to U(\CH_0)$ can be a nontrivial map of $X$ to the
unitary group of $\CH_0$. (It can even be topologically nontrivial, depending on the topology used
to define the unitary group.)

\subsection{Projected Bundles}

Now suppose that $P(x): \CH_x \to \CH_x$ is a continuous family of projection operators. We can then
define a sub-bundle $\CV \subset \CH$ to be the vector bundle whose fiber above $x$ is just the
image of $P(x)$ within $\CH_x$. Then $\CV$ is called the \emph{projected bundle} associated to the
family of projection operators.

There is another useful way to think about projected bundles.
Recall that a \emph{section} of $\pi: \CH\to X $ is a continuous map $\Psi: X \to \CH$ so that
$\pi(\Psi(x)) = x$. That is, a section is a continuous assignment of vectors
\be
x \mapsto \psi(x) \in \CH_x
\ee
(We will generally denote sections by capital Greek letters like $\Psi$ and the values of sections at $x$ by
lowercase Greek letters like $\psi(x)$.)  Now, we can define a projected bundle by saying what its space of
sections is. The linear span of these sections at any $x$ then defines the fiber $\CV_x \subset \CH_x$.
By definition, the   space of sections $\Gamma(\CV)$ is the set of sections
which are eigenvectors of $P(x)$ of eigenvalue $1$ for all $x$:
\be
\Gamma(\CV) := \{ \Psi \in \Gamma(\CH):  P(x) \psi(x) = \psi(x) \qquad \forall x \}.
\ee

An important theorem, the Serre-Swan theorem, says that every finite dimensional vector bundle is isomorphic to a projected
subbundle of some finite rank trivial Hilbert bundle.

\subsection{Projected Connections}

Let us first recall the definition of a connection on a vector bundle.
A connection is simply   a first-order differential operator on sections of $\CV$. That is, one can
define a connection $\nabla$ on $\CV$ to be a map
\be
\nabla: \Gamma(\CV) \to \Omega^1(\CV),
\ee
where $\Omega^1(\CV)$ is the space of one-forms on $X$ valued in $\CV$, such that the
Leibniz rule is satisfied:
\be
\nabla(f \Psi) = df \otimes \Psi + f \nabla(\Psi).
\ee
Here $f$ is an arbitrary differentiable function on $X$. Note that the difference of
two connection is a one-form, valued in endomorphisms of the fibers of $\CV$.
Indeed, the space of connections on a vector bundle $\CV$ is an \underline{affine space} modeled on the vector space
$\Omega^1({\rm End}(\CV))$. It has no natural choice of origin and hence there is no canonical connection on $\CV$.
 There is one important exception to this statement: If the vector bundle $\CV$ is
trivial,   $\CV = X \times \CV_0$ for some fixed vector space $\CV_0$ then there is a natural choice of origin,
namely the trivial connection. To define it, choose any basis $v_n$ of $\CV_0$ so the general section is
$\Psi = s^n v_n$ where $s^n$ is a tuple of functions on $X$. Then the trivial connection is defined by
\be
\nabla^{\rm trivial} \Psi = dx^\mu \frac{\p s_n}{\p x^\mu} \otimes v_n .
\ee

Now suppose that $\CV$ is presented as a projected subbundle of a Hilbert bundle $\pi: \CH \to X$, \underline{and}
we have chosen a connection $\nabla^{\CH}$ on $\CH$. Then we can define the \emph{projected connection}:
\be
\nabla^P := P\circ \nabla^{\CH} \circ \iota
\ee
where $\iota: \Gamma(\CV) \to \Gamma(\CH)$ is the inclusion and $P: \Gamma(\CH) \to \Gamma(\CV)$ is the projection.

 While the above definition of a connection is most convenient for defining
projected connections, there is a more conceptual definition
which will prove useful in section \ref{sec:BandStructure} below.
In this second definition a connection on a vector bundle $\pi: \CV \to X$ is simply a rule for coherently
lifting paths in $X$ to paths in $\CV$.
A ``lifting'' of a path $\gamma:[0,1] \to X$ to a path $\tilde \gamma:[0,1] \to \CV$
is a path $\tilde\gamma$ such that  $\pi(\tilde\gamma(t)) = \gamma(t)$. By ``coherently lifting'' we
mean that the lifted paths satisfy natural composition laws.
\footnote{
To be a little more precise, a connection is a rule which takes a pair of
data given by: (1.)  a continuous path $\gamma:[0,1] \to X$ connecting, say,
$x_0$  to $x_1$ and (2.)
a choice of initial vector $v_0 \in \pi^{-1}(x_0) = \CV_{x_0}$ and returns a lifted
path $\tilde \gamma: [0,1] \to \CV$ with $\tilde\gamma(0)= v_0$.  The lifted path is required to satisfy a
natural composition rule so that composition of paths in the base corresponds to suitable composition of lifted
paths. In equations, if $\gamma_{0,1}$ is a path from $x_0$ to $x_1$ and $\gamma_{1,2}$ is
a path from $x_1$ to $x_2$ and we define $\gamma_{0,2}$ to be the composed path
  $\gamma_{0,1}\circ \gamma_{1,2}$ (with time running twice as fast) then if the choice of
  initial vector $v_0 \in \CV_{x_0}$ defines a lift with endpoint $v_1 = \tilde\gamma_{0,1}(1)$
  and we use $v_1$ as the initial vector for lifting $\gamma_{1,2}$ then the composition
$\tilde\gamma_{0,1}\circ \tilde \gamma_{1,2}$ is the lift of $\gamma_{0,2}$ with initial vector
$v_0$. Finally, the lifting must be linear: If the initial points $v_0$ and $v_0'$
lead to lifted paths $\tilde\gamma$  and $\tilde\gamma'$ then the initial point
$v_0'' = v_0 + v_0'$ leads to a lifted path with $\tilde\gamma''(t) = \tilde\gamma(t) +
\tilde\gamma'(t)$. The addition here means vector addition in the fiber above $\gamma(t)$. }
 The relation between the two definitions
is this: To define a lifted path $\tilde\gamma(t)$ one solves the first order differential equation $\nabla_{\frac{\p}{\p t}} \Psi =0$,
where $\frac{\p}{\p t}$ is the tangent vector to $\gamma(t)$,
using the initial condition $\tilde\gamma(0)=v_0$. It must be stressed that the projection of the lifted path in $\CH$ from
$\nabla^{\CH}$ is in general \underline{not} the lifted path in $\CV$ from $\nabla^P$! Indeed, even in the
case that  $\CH$ is
the trivial bundle and we use the trivial connection $\nabla^{\CH} = \nabla^{\rm trivial}$
the lifted path in $\CH$ is $\tilde \gamma(t) = ( \gamma(t), v_0)$
where $v_0$ is the initial vector in the fiber over $\CH_{\gamma(0)}$.
However the projected path $(\gamma(t), P(\gamma(t))v_0)$ will not
be covariantly constant with respect to $\nabla^P$ precisely because, in general, $PdP \not= 0$. In this way
the projected connection can be nontrivial, even if $\nabla^{\CH}$ is trivial.

A simple and popular example of a projected connection is given by choosing the Hilbert bundle
\be
\CH = S^2 \times \IC^2
\ee
where we think of $S^2$ as the set of unit vectors in $\IR^3$ and we choose the family of projection
operators
\be
P(\hat x) = \half (1 + \hat x \cdot \vec \sigma).
\ee
In this case, if we take the trivial connection on $\CH$ then the projected connection is just the
magnetic monopole connection on a complex line bundle of first Chern class one.

We are finally ready to define Berry connections. Suppose then we have a Hilbert bundle $\pi: \CH \to X$
and moreover we are given a  continuously varying family of Hamiltonians $H_x$ acting on $\CH_x$ so that there is an energy
$E_{\rm gap}$ that is not in the joint spectrum of all the Hamiltonians:
\be
E_{\rm gap} \notin \cup_{x\in X} {\rm Spec}(H_x).
\ee
In this case we can define projection operators $P(x)$ to project onto the eigenstates of $H_x$ of energies
below $E_{\rm gap}$:
\be
P(x) = \Theta( E_{\rm gap} - H_x)
\ee
where $\Theta$ is the Heaviside step function. If we \underline{choose} a connection $\nabla^{\CH}$ on $\CH$
then the projected connection is known, in physics, as a Berry connection:
\be
\nabla^{\rm Berry} :=  P \circ \nabla^{\CH} \circ \iota .
\ee

The central point we are attempting to make in this
note is simply to stress that the definition of the Berry connection depends on a choice of connection
$\nabla^{\CH}$ on $\CH$. In general the Hilbert bundle $\CH$ is trivializable, but there is no natural
trivializtion.   When this is the case there is no natural origin in the affine space of connections on
$\CH$. In the original papers by M. Berry \cite{Berry1}\cite{Berry2} and by B. Simon \cite{Simon}
it was assumed that $\CH$ is the trivial bundle. In this case, as we have noted, there is a natural
origin, namely the trivial connection on $\CH$ and that is the connection implicitly used in those papers
(and all subsequent papers of which we are aware). Thus, when one encounters expressions like
\be\label{eq:StandardBerry}
\vec A^{\rm Berry} = \langle \psi \vert \vec \nabla_{\vec R} \vert \psi \rangle,
\ee
a trivialization has been assumed.
Of course, the expression \eqref{eq:StandardBerry} has been used without incident in a wide variety of applications in molecular,
and optical physics \cite{BohmBook,Shapere:1989kp}. The reason is that in the models used in these applications
the statement of the problem includes a natural trivialization. For example, in the Born-Oppenheimer
approximation in molecular physics the definition of the Hilbert space of the electrons does not
depend on the positions of the nuclei (although the Hamiltonian most certainly does). Therefore
the original Hilbert bundle over the control parameter space is literally a trivial bundle.

\section{The Adiabatic Theorem Revisited}

The Berry connection has its origin in the quantum adiabatic theorem \cite{Berry1,Simon}. In this case we study the
Schr\"odinger equation
\be\label{eq:Schrod}
\I \hbar \frac{\p \Psi}{\p t} = H(X(t)) \Psi
\ee
where the $X(t)$ are time-dependent control parameters, varying slowly on the time-scales set by
the energies of the Hamiltonians $H(X(t))=: H(t)$. How do our current considerations enter into the
adiabatic theorem? Although universally written (so far as we know), equation \eqref{eq:Schrod} is actually
not completely sensible from a mathematical viewpoint: If the $X(t)$ are
varying then $\Psi(t)$ must be a section of a bundle of Hilbert spaces and in particular,
for each different $X(t)$ the vector $\Psi(t) \in \CH_{X(t)}$ is in a different fiber of the Hilbert bundle.
 If the bundle has not been trivialized
then it does not make sense to write a derivative $\frac{\p \Psi}{\p t}$ without a choice of
connection on the bundle!  The correct Schr\"odinger equation must actually be  written as:
\be\label{eq:Schrod-Conn}
\I \hbar \nabla^{\CH}_{\frac{\p}{\p t}} \Psi  = H(X(t)) \Psi
\ee
where  $\nabla^{\CH}_{\frac{\p}{\p t}}$ is a contraction of the connection $\nabla^{\CH}$ on $\CH$
with the tangent vector to the path in the space of control parameters.

Now recall that the essence of the  adiabatic theorem for \eqref{eq:Schrod} is that
the natural parallel transport defined by the time-evolution operator
$U(t_1, t_2) = {\rm Pexp} \frac{1}{\I \hbar} \int_{t_2}^{t_1} H(t') dt'$
does not commute with the spectral projections for the Hamiltonians $H(t_1)$ and $H(t_2)$.  However, if the
variation of $H(t)$ is ``small'' then the time-evolution operator has a ``small'' commutator
with the spectral projections: The amplitudes for transitions between eigenspaces for different
eigenvalues
will be small. The same general discussion can be applied to \eqref{eq:Schrod-Conn} and of
course the connection $\nabla^{\CH}$ must likewise be such that the associated parallel
transport operators have ``small'' commutators with the spectral projections of $H(t)$.
When this holds the   quantum adiabatic theorem can be
generalized, and after removing the ``dynamical phase''
(for example, the factor $\exp[ \frac{1}{\I \hbar} \int E_n(t') dt']$ associated with
an isolated eigenvalue $E_n(t)$) the remaining ``geometric phase'' is just the parallel
transport operator associated with the projected connection for $\nabla^{\CH}$.

\section{Band Structure}\label{sec:BandStructure}

A simple example of a Hilbert bundle with no natural trivialization is provided by the familiar
setting of band structure. Let $C \subset \IA^n$ be a crystal in affine $n$-dimensional space
with Euclidean metric. We assume that $C$ is invariant under translation by a rank $n$ lattice
$L \subset \IR^n$. We can then define the Brillouin torus $T^\vee$ to be the space of unitary
irreducible representations of $L$. This is generally presented as $T^\vee = \CK/L^\vee$
where $L^\vee:=\Hom(L,\IZ)$ is the integral dual of $L$ and $\CK$ is the vector space of momentum vectors.
$\CK$ is also known as ``reciprocal space.''
\footnote{Here we differ from traditional solid state physics conventions where the inner product
between a lattice and reciprocal lattice vector is $2\pi$ times an integer. In our conventions
the inner product is integral. Consequently, there is
an extra $2\pi$ in the formula for the Bloch phases below relative to what is standard in
the condensed matter literature. }
 Mathematically, it
can be identified with the cotangent space of $\IA^n$ at any point. We denote points of the Brillouin torus
by $\bar k$ so that the corresponding unitary  representation of $L$ can be presented as
\be
\chi_{\bar k}(R) = e^{2\pi \I k \cdot R }
\ee
for all $R\in L$, where $k$ is any lift of $\bar k$ to $\CK$.

We can define a Hilbert bundle
$\pi: \CH \to T^\vee$ whose fiber $\CH_{\bar k}$ is the space of Bloch functions on $\IA^n$:
\be\label{eq:Bloch}
\psi(x+R) = e^{2\pi \I k \cdot R} \psi(x)
\ee
that are suitably normalizable. (They are $L^2$-normalizable over a fundamental domain for
the action of $L$ on $\IA^n$.) Put differently, $\CH_{\bar k}$ is the space of $L^2$-sections of
the Poincar\'e line bundle $L_{\bar k}$ over the torus $T:=\IA^n/L$.

Now, there is no natural trivialization of $\CH$ simply because the definition of the fibers $\CH_{\bar k}$
depends on $\bar k$. In discussions of band structure we are given a Schr\"odinger operator on $L^2(\IA^n)$ invariant under translation
by $L$ and then, under some circumstances, the eigenfunctions $\{ \psi_{n,\bar k} \}$ of the Schr\"odinger operator define
a natural trivialization. Roughly speaking, below any finite energy the bands should not cross and should define
a trivializable finite rank vector bundle. (In particular all the Chern classes and $K$-theory invariants should be trivial.)
In general,
 bands do cross, and Chern numbers are nonzero. In such cases, there is certainly no natural trivialization of the Hilbert bundle.

Suppose now that we have an insulator, so that the Fermi energy $E_f$ of the many-body ground state
is in a gap separating filled (valence) bands from the
unfilled (conduction) bands. This is a classic example of the situation in which we would like to define the Berry connection
associated with the Hamiltonians $H_{\bar k}$ acting on the Bloch functions with Bloch momentum $\bar k$.
 In this case we have a natural projected bundle, $\CF$, the bundle of filled bands, defined
by using $P(\bar k) = \Theta(E_f - H_{\bar k}) $. However, as we just stressed, the Hilbert bundle has no natural trivialization,
therefore, it would seem that  there is no natural choice for $\nabla^{\CH}$ and hence there is no natural choice of the Berry connection.
The situation is not quite that dire because, as noted in \cite{Freed:2012uu}, there is in fact a natural \underline{family} of
connections $\nabla^{\CH,\bar x_0}$ whose gauge equivalence class is parametrized by the real-space torus: $\bar x_0 \in \IA^n/L$.
These connections have the distinguishing property that they are flat (zero curvature) and their holonomy only depends on the data at hand.
They also show up in first order response theory.

We recall here the definition of $\nabla^{\CH,\bar x_0}$ from \cite{Freed:2012uu}. Since the connections are all flat connections
it suffices to define the parallel transport on straight-line paths and these can in turn be taken to be projections of paths in
$\CK$ given by
\be\label{eq:k-path}
k(t) = k_0 + t \delta k  \qquad 0 \leq t \leq 1.
\ee
Let $\bar \gamma(t)= \overline{k(t)}$ denote the projection of the path \eqref{eq:k-path} in $\CK$ to a path in $T^\vee$.
We need to define a lift of $\bar\gamma(t)$ to a path $\tilde \gamma(t)$ in $\CH$,
given an initial vector $\psi_{\bar k_0} \in \CH_{\bar k_0}$. If we choose $x_0 \in \IA^n$ then we can define
$\tilde\gamma(t) = U^{\CH,x_0}(t) \psi_{\bar k_0}$ to be the function on $\IA^n$ whose value at $x$ is
\be
(U^{\CH,x_0}(t)\cdot \psi_{\bar k_0})(x):= e^{2\pi \I t \delta k\cdot(x - x_0)} \psi_{\bar k_0}(x).
\ee
It is easily checked that $U^{\CH,x_0}(t) \psi_{\bar k_0}$ has quasi-periodicity given by $\overline{k(t)}$  and that
paths suitably compose. By computing the holonomy around small closed triangles one also easily checks that this is
a flat connection.  Note especially that since the crystal is in affine space we must subtract points $x-x_0$ in
order to define a vector which can be paired with a reciprocal vector $\delta k \in \CK$. It would not make sense to write
this formula without $x_0$. If you do that, you are implicitly making an unphysical choice of origin.
Thus, we need to choose
a point $x_0$ in order to define the connection. The isomorphism class (= gauge equivalence class) of a flat connection is
completely characterized by its holonomy. Using $\pi_1(T^\vee, \bar k_0) \cong L^\vee$ for each homotopy class we can choose the
minimal length representative to be the projection of
\be
k(t) = k_0 + t G \qquad 0 \leq t \leq 1
\ee
where $G\in L^\vee$ is a reciprocal lattice vector.
Denote the corresponding closed curve in $T^\vee$ by  $\gamma_G$. (The notation suppresses the dependence
on the basepoint $\bar k_0$.)  The holonomy around  $\gamma_G$ is
the multiplication of Bloch functions:
\be
H^{x_0}(\gamma_G) : \psi_{\bar k_0}(x) \mapsto e^{2\pi \I G\cdot (x-x_0)} \psi_{\bar k_0}(x).
\ee
The difference of holonomy operators for two choices of $x_0$ is
\be
H^{x_0}(\gamma_G)( H^{x_0'}(\gamma_G) )^{-1} = e^{-2\pi \I G\cdot (x_0 -x_0')}\textbf{1}_{\CH_{\bar k_0}}
\ee
and is proportional to the unit operator on $\CH_{\bar k_0}$. Note that $H^{x_0}(\gamma_G) $
 only depends on the projection of $x_0$ to $T$.
Thus the isomorphism classes of the connections $\nabla^{\CH,x_0}$ are parametrized by the real-space torus $T$.

Another way to say this is that if we view the connection $\nabla^{\CH,x_0}$ as a first order differential operator
on $\Gamma(\CH)$ then the difference of two connections in the family is:
\be
\nabla^{\CH,x_0} - \nabla^{\CH,x_0'} = 2\pi \I dk \cdot (x_0 -x_0') \otimes \textbf{1}_{\CH}.
\ee
A shift of $x_0$ by a lattice vector $R\in L$ can be undone by a gauge transformation defined by
conjugating all Bloch functions in $\CH_{\bar k}$ by the unitary operator $U(\bar k) = e^{2\pi \I k \cdot R}$.
Thus, we see again that the gauge equivalence class of $\nabla^{\CH,x_0}$ only depends on the projection
of $x_0 $ to $\bar x_0 \in T$. If we now consider an insulator with a bundle of filled bands $\CF$ then the projected connections
associated to a Fermi energy $E_f$ together with a  choice $\nabla^{\CH, \bar x_0}$ on $\CH$ gives us a corresponding
family of Berry connections $\nabla^{\CF, x_0}$ such that
\be\label{eq:BerryFamily}
\nabla^{\CF,x_0} - \nabla^{\CF,x_0'} = 2\pi \I dk \cdot (x_0 -x_0') \otimes \textbf{1}_{\CF}.
\ee
Again, the gauge equivalence classes of these connections are parametrized by the real space torus $T$.
 If one thinks of the Berry connection in band theory as just given by the
 expectation value of the position operator then the dependence on a choice of origin is completely obvious.

The subtlety we are stressing here can often, but not always, be ignored.
It is important to distinguish a crystal $C \subset \IA^n$ in an affine space with
no distinguished origin from a lattice $L \subset \IR^n$, which does have a distinguished origin. If $L$
acts transitively on $C$ (that is, there is just one atom per unit cell) then there \underline{is} a natural
choice for $\bar x_0$, provided by the equivalence class of the points $C$ itself. In this case there is a
 distinguished Berry connection. Moreover,   the difference of the connections in \eqref{eq:BerryFamily}
 is a flat one-form proportional to the unit operator on the fibers of
$\CF$. Thus the difference might seem a bit trivial, and indeed the field strengths $F(\nabla^{\CF,x_0})$ are
independent of $x_0$. Therefore the Chern-Weil representatives of Chern classes
 are independent of $x_0$. (This is in harmony with the
fact that the Chern classes of $\CF$ cannot depend on any choice of connection on $\CF$.) Nevertheless,
in general, the gauge equivalence class of a connection on a vector bundle is \underline{not} completely captured
by the fieldstrength. Two classic examples of gauge invariant information not captured by the
fieldstrength are holonomy around nonbounding closed cycles and Chern-Simons invariants. Both of these examples
show up in the physics of insulators. We will now discuss these two examples.

\section{Example 1: Electric Polarization}

There is a famous formula for the contribution of the valence electrons in an insulator to the
zero-temperature electric polarization per unit volume, $P$. It must be admitted that, in fact,
$P$ is not quite well-defined. Only differences of $P$ are really well-defined. Nevertheless
there is a useful expression for $P$, defined up to shifts by $\frac{e}{\Omega} L$, where $\Omega$ is the
volume of a unit cell of $L$, from which one can derive the physically relevant differences of
$P$ resulting from a change in control parameters.  The ``momentum space'' expression
(equation $(8b)$ of  \cite{KingSmithVanderbilt} or  equation $(19)$ of  \cite{RestaVanderbilt}) can be written
in the form:
\be\label{eq:BerryPol}
G \cdot P = \frac{e}{2\pi\I}   \int_{T^{\vee,\perp}} \log \, \det \, {\rm Hol}(\nabla^{\CF,x_0}(\gamma_G))~{\rm d\, vol}  ~~~ \mod ~ \frac{e}{\Omega}\IZ
\ee
where $G$ is an arbitrary reciprocal lattice vector and $\gamma_G$ is the corresponding closed loop in $T^\vee$.
We have written this in terms of the holonomy of the Berry connection, ${\rm Hol}(\nabla^{\CF,x_0}(\gamma_G))$
to emphasize the gauge invariance of the result. $T^{\vee,\perp}$ is the subtorus through the origin
\footnote{In general $\gamma_G$ depends on a choice of basepoint $\bar k_0$, and if we choose $\bar k_0 \not=0$ we should choose
the subtorus through that point. More on that below.}
and orthogonal to $G$ and it carries
a natural volume form ${\rm d \, vol} $ inherited from the Euclidean metric on $\IA^3$.
The expression \eqref{eq:BerryPol} involves taking a logarithm of the gauge-invariant
holonomy of the Berry connection, and consequently $P$ is only defined modulo $\frac{e}{\Omega}$ times a lattice vector.
The ambiguity of $P$ under shifts in $\frac{e}{\Omega} L$
 is well-appreciated and widely discussed in the literature, and the physical origin of this
ambituity is well-understood. See, for example
\cite{KingSmithVanderbilt,RestaVanderbilt,RestaRev}. On the other hand, the expression also depends
\underline{continuously} on $x_0$ since:
\be\label{eq:OriginDep}
P(x_0) - P(x_0') = \frac{e}{\Omega} (x_0 - x_0').
\ee
This is completely distinct from the usual ambiguity of the polarization by $\frac{e}{\Omega}  L$  discussed in the literature.
On the other hand, the ambiguity \eqref{eq:OriginDep},  is also physically rather trivial: It is exactly what we expect for the dependence
of electric polarization on a choice of origin if we only take into account the contribution of
the electrons - all of which have the same sign! The physically relevant result must take into account
the contribution of the neutralizing positive charges of the ions as in equation $(20)$ of
\cite{RestaVanderbilt}.

\bigskip
\noindent
\textbf{Remarks}

\begin{enumerate}

\item The derivation of \eqref{eq:BerryPol} begins with a Kubo formula for the variation
of polarization as a function of control parameters in some space $X$ \cite{Resta}. The
Kubo formula was used to define  a one-form $dP$ on the product $T^\vee \times X$.
It was observed in \cite{Rabe} that this one-form in fact has nonzero periods, and a relation
to Berry's phase was noted.  In
\cite{KingSmithVanderbilt} it was observed that a local anti-derivative of the one-form can be
written in a form equivalent to \eqref{eq:BerryPol}, and it is in this step, where one passes from
Berry curvature to holonomy,  that the $x_0$ dependence enters.

\item  In some cases when one has a global trivialization $\{ \psi_{n,\bar k} \}$ of $\CH$
 one can introduce a set of ``Wannier functions'':
\be\label{eq:Wannier}
W_{n,R} : = \frac{1}{(2\pi)^3} \int_{T^\vee} \chi_{\bar k}^*(R) \psi_{n,\bar k} ~{\rm d\, vol}
\ee
In such cases there is  an alternative ``real space'' expression (see equation $(10)$ of
\cite{KingSmithVanderbilt}).  These real-space expressions make the dependence of $P$ on a choice of
origin obvious, and we have now explained where that dependence resides in the reciprocal
space version of the formula, namely equation \eqref{eq:BerryPol}. (We also note in passing that
\eqref{eq:Wannier} only makes sense when we can choose a global trivialization, and then only
because $\psi_{n,\bar k}$ can be viewed as a quasiperiodic function on $\IA^n$ and it makes sense to
add such functions.  In general it   does not
make mathematical sense to add vectors in different fibers of a bundle. Rather, one must
use a connection to parallel transport these vectors to a common fiber, where they can be
added. Thus, in general, one might attempt to choose a basepoint  $\bar k_0\in T^\vee$, and
a flat connection, such as $\nabla^{\CH, x_0}$ and a choice of paths
$\gamma_{\bar k, \bar k_0}$ from $\bar k \in T^\vee $ to $\bar k_0$ and write
\be\label{eq:WannierCorrect}
W_{n,R} : = \frac{1}{(2\pi)^3} \int_{T^\vee} \chi_{\bar k}^*(R) U^{\CH,x_0}(\gamma_{\bar k, \bar k_0} )\psi_{n,\bar k}
~{\rm d\, vol}
\ee
A natural way to choose paths is to choose a fundamental domain for $L^\vee$ in $\CK$ and use straight-line paths
from $\bar k_0$ to the boundaries. Nevertheless, this is not totally satisfactory since the result will still depend on
the choice of fundamental domain.)

\item In the case of a Chern insulator, where the Berry curvature has periods, the expression \eqref{eq:BerryPol}
has further dependence on the basepoint of the closed loop $\gamma_G$. Thus it is necessary to specify
further data to obtain a meaningful expression for $P$ \cite{CohVanderbilt}.

\end{enumerate}

\section{Example 2: Axion Angle}

As a second example we consider the case of the magneto-electric polarizability response tensor
for an insulator in $3+1$ dimensions. This can be defined as the leading term in the low energy effective action for the
Maxwell gauge field in the presence of an insulator:
\be
\CS_{\rm eff} \sim \frac{1}{\hbar} \int_{\IR^4} \alpha^{ij} E_i B_j + \cdots
\ee
where $E_i$ and $B_j$ are the components of the electric and magnetic fields of an external Maxwell gauge field
probing the insulator. In a rather beautiful development \cite{Essin:2008rq,Essin:2010in,Qi:2008ew}, it was noted
that the trace part of $\alpha^{ij}$ (to be thought of as an ``axion angle'') is related to the Chern-Simons invariant
of ``the'' Berry connection on the Brillouin torus. Since there is, in general, no distinguished Berry
connection we must choose one, and the ones which arise from first order response theory come from the
family of connections $\nabla^{\CF,x_0}$ described above. So we
write the theta angle in terms of the Chern-Simons invariant of this connection:
\be
\theta(x_0) = \frac{1}{3} \alpha^{ii} = \int_{T^\vee} {\rm CS}(\nabla^{\CF,x_0}).
\ee
Our normalization of the Chern-Simons invariant is that
$\theta$ is defined modulo $2\pi$. In \cite{Essin:2008rq,Essin:2010in,Qi:2008ew} the $x_0$ dependence was not included.
In the case of insulators with time-reversal invariance or parity invariance the connection must be compatible with the
lift of $P$ or $T$ to $\CF$ and, as is well-known, the Chern-Simons invariant is either $0$ or $\pi$ modulo $2\pi$.
 Indeed, one can define a strong topological insulator as a parity or
time-reversal invariant insulator with $\theta = \pi$. However, for a general insulator there is no natural
choice of basepoint $x_0$. Quite generally, under a shift of connection $\nabla \to \nabla + \alpha$, where
$\alpha$ is a one-form valued in endomorphisms of the vector bundle, the Chern-Simons
form changes by
\be
{\rm CS}(\nabla+ \alpha) - {\rm CS}(\nabla) = {\rm Tr} \left( 2 \alpha F + \alpha \nabla \alpha + \frac{2}{3} \alpha^3 \right).
\ee
We use this equation with $\alpha = 2\pi \I dk\cdot (x_0'-x_0) \textbf{1}_{\CF}$ as in \eqref{eq:BerryFamily} to obtain
\be
{\rm CS}(\nabla^{\CF,x_0'}) = {\rm CS}(\nabla^{\CF,x_0})  +  4\pi \I  dk\cdot (x_0'-x_0) \wedge {\rm Tr}   F(\nabla^{\CF,x_0}).
\ee
In our conventions,  ${\rm Tr} F(\nabla^{\CF,x_0})$ is $\frac{\I}{2}$ times
 the Chern-Weil representative of the first Chern class  $c_1(\CF)$  of $\CF$. The first Chern class $c_1(\CF)$
can be integrated over a basis of homology two-cycles in $T^\vee$ to produce a three-component vector.
In fact this vector can be understood more invariantly by interpreting the expression
\be
\int_{T^\vee} c_1(\CF)
\ee
as a vector in the reciprocal lattice $L^\vee$ as follows:
Recall that $H_1(T^\vee;\IZ) \cong L^\vee$. Therefore $H^1(T^\vee;\IZ) \cong L$
and hence $H^2(T^\vee;\IZ) \cong L \wedge L $.
Therefore we can interpret the map $H^1(T^\vee;\IZ) \to \IZ$ defined by
\be
\omega \mapsto \int_{T^\vee} \omega \wedge c_1(\CF) \in \IZ
\ee
as:
\be
L ~  {\buildrel \wedge c_1(\CF) \over \longrightarrow } ~ L \wedge L \wedge L  ~   \cong \IZ
\ee
where in the second step we paired with the fundamental class of $T^\vee$.
(Of course, one can also evaluate the integral directly using
local expressions to arrive at the same result.) We will denote the resulting reciprocal lattice vector
 by $ G_{\rm CH}$ and refer to it as the \emph{Chern-Halperin vector}.
The net result is that the dependence of the axion angle on origin is simply:
\be\label{eq:AxionDep}
\theta(x_0) - \theta(x_0') = 2\pi G_{\rm CH} \cdot (x_0 - x_0') \mod 2\pi .
\ee

How should we interpret the dependence on $x_0$? One natural way is to interpret the statement as saying
that for a three-dimensional Chern insulator the axion angle must be linear in space with proportionality
constant given by the Chern-Halperin vector.
\footnote{I thank Edward Witten for an essential discussion about this point.}
 The axion angle is periodic, so in this effective theory a
remnant of the lattice periodicity remains in the periodicity under shifts of $x$ by a lattice vector.
With this interpretation we can integrate by parts so that the axion
angle term in the effective action for the Maxwell gauge field leads to a factor in the path integral:
\be\label{eq:3D-QHE}
\exp\left[ \frac{\I}{4\pi }\left( \frac{e}{\hbar c} \right)^2
\int_{\IR^4}  ( G_{\rm CH} \cdot d x)\wedge A  \wedge dA \right]
\ee
where we have expressed the result in cgs (Gaussian) units and $A$ is the gauge potential for the
Maxwell fieldstrength.  Differentiating with respect to $A_i$ gives the contribution
to the current:
\be
J^i  =  \frac{e^2}{h} \epsilon^{ijk} (G_{\rm CH})_j E_k
\ee
where again we stress that we are using cgs units and $(G_{\rm CH})_j$ is a reciprocal vector to
the lattice vectors so that $G_{\rm CH} \cdot R \in \IZ$ for all $R\in  L $.
This is the three-dimensional quantum Hall effect of  \cite{Halperin1,Halperin2}.

\subsection{A Pseudo-Topological Field Theory}

The dimensionless action for the electromagnetic field (in cgs units) is
\footnote{We are adding the Maxwell action in free space. It would make more sense to
include a general permeability tensor in the first term to take into account the
electromagnetic response of the
material. However, to keep things simple we will ignore this point.}
\be\label{eq:3DQHE-LEET}
\begin{split}
S & = \frac{1}{\hbar c} \int \frac{1}{8\pi} F\wedge *F -  \left( \frac{e}{\hbar c} \right)^2 \int  \frac{ G_{CH} \cdot d  x}{4\pi} \wedge A \wedge F\\
&
 = \frac{1}{8 \pi \hbar c} \int \left( F\wedge * F - G \wedge A \wedge F \right)\\
\end{split}
\ee
where we have defined a   closed one-form on spacetime $G$ given by
$G := \frac{2e^2}{\hbar c}  G_{CH} \cdot d  x$ in  the rest-frame of the
crystal. By power counting of momenta the action \eqref{eq:3D-QHE} should give the leading
long-distance physics. However,  the action
\be
\int G \wedge A \wedge d A
\ee
is singular, even after allowing for gauge invariance: There is a
zero mode under $A \to A + \lambda G$, where $\lambda$ is any real number.
The situation is a little reminiscent of Costello's discussion of integrable
lattice models \cite{Costello:2013sla}. In order to have a nonsingular theory
we must include the Maxwell action and hence the $3+1$ dimensional theory is
only ``partially topological.'' In momentum space the action is
\be
S = \frac{1}{8 \pi \hbar c}  \int \frac{d^4 k}{(2\pi)^4} A^\mu(-k) \CD_{\mu\nu} A^\nu(k)
\ee
with
\be
\CD_{\mu\nu} = ( \eta_{\mu\nu} k^2 - k_\mu k_\nu) + \I \epsilon_{\mu\nu\lambda\rho} G^\lambda k^\rho
\ee
  We can only hope to
invert $\CD_{\mu\nu}$ on a space orthogonal to the gauge modes, so we look for $P^{\nu\lambda}$ such that
\be\label{eq:Invert}
\CD_{\mu\nu} P^{\nu\lambda} = \delta_\mu^{~\lambda} -  \frac{ k_\mu k^\lambda}{k^2}.
\ee
Now $P^{\nu\lambda}$ must be a linear combination of the tensors:
\be
P^{\nu\lambda}= F_1  \eta^{\nu\lambda} + F_2 k^\nu k^\lambda + F_3 k^\nu G^\lambda + F_4 G^\nu k^\lambda
+ F_5 G^\nu G^\lambda + F_6 \epsilon^{\nu\lambda\zeta\omega} G_{\zeta} k_{\omega}
\ee
where $F_1, \dots , F_6$ are functions of $k^2, G^2,(G\cdot k)$. Imposing the condition
\eqref{eq:Invert} the functions $F_2$
and $F_3$ are undetermined while the remaining functions are uniquely determined.  The function $F_3$
is determined by requiring $P^{\nu\lambda}(k)$ to be
symmetric under $\nu \leftrightarrow \lambda$ and $k \to -k$ and the choice of
$F_2$ is a choice of gauge. Putting $F_2=0$ we get:
\be
-\I P^{\nu\lambda} = \frac{ k^2 \eta^{\nu\lambda} -   \frac{G\cdot k}{k^2} (G^\nu k^\lambda + k^\nu G^\lambda)
 +   G^\nu G^\lambda
-\I  \epsilon^{\nu\lambda \zeta \omega} G_{\zeta} k_{\omega} }{G^2 k^2 - (G\cdot k)^2 + k^4 }
\ee
In Euclidean signature the denominator does not vanish, by the Cauchy-Schwarz inequality.

If we count derivatives in the action \eqref{eq:3DQHE-LEET} then we might naively expect
the Chern-Simons term to dominate in the infrared ($k\to 0$) limit. This is not quite true
since the numerator of the propagator has terms of order zero in the $k$-expansion. But if we consider polarizations
orthogonal to $G$ then indeed as $k\to 0$ we find a propagator scaling like $1/k$,
similar to the standard propagator in Chern-Simons perturbation theory:
\be
P^{\nu\lambda}  \sim \frac{ \epsilon^{\nu\lambda \zeta \omega} G_{\zeta} k_{\omega} }{G^2 k^2 - (G\cdot k)^2 }
\ee
so long as we restrict attention to momenta $k$ not parallel to $G$. Note the singularity when $k \parallel G$.
Like Hamlet, who is but mad north-north-west, the $k\to 0$ limit of the theory is only non-topological in the
direction (in momentum space) parallel to $G$.

\subsection{Dislocations}

It was already noted in \cite{Halperin1} that in the presence of dislocations of the
three-dimensional crystal there will be interesting ``edge modes'' localized at those
dislocations. A particularly obvious case is that of a screw dislocation along an
axis (say, the $z$-axis) of a stack of two-dimensional Chern insulators. In this case,
if we cut a small hole out around the dislocation with a helical boundary then the
familiar  edge modes will be localized on that helix.
  A general approach to the theory of such modes can be found in
\cite{Bulmash}. A related result for helical modes in topological insulators
can be found in \cite{Hughes:2013xaa,RanDislocation}. Here we discuss the subject using the
derivation of edge modes explained, for example, in \cite{Elitzur:1989nr,Moore:1989yh,Moore:1989vd,Witten:1988hf}. It just
implements the anomaly inflow mechanism of Callan and Harvey \cite{Callan:1984sa}.
The Chern-Simons term is anomalous under singular gauge transformations of electromagnetism, singular
on the dislocation line. The anomaly must be cancelled by modes on a ``cosmic string''
or, in more modern language, on a ``surface defect'' in the effective gauge theory.
\footnote{This terminology could potentially be confusing: What high energy
physicists call ``surface defects'' would be called ``line defects'' in
condensed matter theory. We will deprecate the term ``line defects''
(in the condensed matter sense) in favor of ``dislocation lines.''}

In the normal plane to a dislocation line the lattice has a locally-defined infinitesimal
displacement by a vector $ u(\phi)$ (where $\phi$ is an angular coordinate in
the normal bundle) such that
\be
\oint d \phi \frac{d u}{d \phi} :=  D \in L
\ee
where $ D$ is known as a \emph{Burgers vector}. Thanks to \eqref{eq:AxionDep}
the axion angle has the property that
\be
\oint d\theta =2\pi  G_{CH} \cdot  D.
\ee
Now, a singular gauge transformation for the $U(1)$ electromagnetic field  $g(\phi) =e^{\I \phi}$ leads to
an anomalous change in the action by
\be
( G_{CH} \cdot  D ) \int_{\Sigma} F
\ee
on the worldsurface $\Sigma$ of the surface defect. This can be canceled by $G_{CH}\cdot D$ chiral modes
located on the surface defect, thus confirming the general arguments of
\cite{Bulmash,Halperin1}.

\subsection{Speculation: A $3+1$-Dimensional FQHE}

The $3+1$ dimensional QHE is often dismissed as   ``uninteresting'' because there is a natural layer structure given by
the planes in the crystal orthogonal to $G$ and, it is claimed, it is ``just'' a stack of 2d Chern insulators, and there
will be no phase transition as interlayer couplings are increased. However, having expressed the effective theory of the
$3+1$ QHE in the form \eqref{eq:3DQHE-LEET} there is an obvious generalization, given the well-known formulation of the $2+1$ dimensional
FQHE in terms of abelian spin Chern-Simons theory
\cite{Belov:2005ze,Wen,Frohlich:1991wb,Frohlich:1994nk,Read,Zee}.  We choose a symmetric matrix of
one-forms $K_{IJ}$ where each matrix element is defined by reciprocal lattice vectors
 and introduce an abelian gauge theory for a torus,
with gauge fields $a^I$ (relative to a basis for the Lie algebra of the torus) and finally we choose a
vector of one-forms defined by reciprocal lattice vectors $Q_I$ and consider the action:
\be
\int K_{IJ} \wedge  a^I \wedge d a^J  +   Q_I \wedge da^I \wedge A
\ee
 It would be amusing to
explore the physical implications of such a phase of matter, and whether it can be realized
by interacting electrons in $3+1$ dimensions. In general, if the $K_{IJ}$ do not all point in the
same direction there is no reason why this should behave like a stack of FQHE systems.

\end{document}